\documentclass[
 aip, apl, amsmath, amssymb, reprint]{revtex4-1}
\usepackage{graphicx}
\usepackage{dcolumn}
\usepackage{amsmath}
\usepackage{bm}
\usepackage{subcaption}
\usepackage{multirow}
\usepackage[justification=raggedright]{caption}
\usepackage{float}
\usepackage[demo,abs]{overpic}
\usepackage{color,soul}
 \usepackage{float}
\makeatletter
\let\newfloat\newfloat@ltx
\makeatother
\usepackage{algorithm}
\usepackage{algpseudocode}
\usepackage{booktabs}
\usepackage{color,soul}
\usepackage{tabularx}

\begin{document}

\preprint{AIP/123-QED}

\title{Gradient descent optimization of acoustic holograms for transcranial focused ultrasound}

\author{Ahmed Sallam}
\author{Ceren Cengiz}
\author{Mihir Pewekar}
\author{Eric Hoffmann}
\affiliation{Department of Mechanical Engineering, Virginia Tech, Blacksburg, VA, $24061$, USA}
\author{Wynn Legon}
\affiliation{Fralin Biomedical Research Institute, Virginia Tech, Roanoke, VA, $24016$, USA}
\author{Eli Vlaisavljevich}
\affiliation{Department of Biomedical Engineering and Mechanics, Virginia Tech, Blacksburg, VA, $24061$, USA}
\author{Shima Shahab}
\email{sshahab@vt.edu}
\affiliation{Department of Mechanical Engineering, Virginia Tech, Blacksburg, VA, $24061$, USA}

\begin{abstract}

Acoustic holographic lenses, also known as acoustic holograms, can change the phase of a transmitted wavefront in order to shape and construct complex ultrasound pressure fields, often for focusing the acoustic energy on a target region. These lenses have been proposed for transcranial focused ultrasound (tFUS) to create diffraction-limited focal zones that target specific brain regions while compensating for skull aberration. Holograms for tFUS are currently designed using time-reversal approaches in full-wave time-domain numerical simulations.  However, such simulations need time-consuming computations, which severely limits the adoption of iterative optimization strategies. Furthermore, in the time-reversal method, the number and distribution of virtual sources can significantly influence the final sound field. Because of the computational constraints, predicting these effects and determining the optimal arrangement is challenging. This study introduces an efficient method for designing acoustic holograms using a volumetric holographic technique to generate focused fields inside the skull.  The proposed method combines a modified mixed-domain method for ultrasonic propagation with a gradient descent iterative optimization algorithm. This approach enables substantially faster holographic computation than previously reported techniques. The iterative process uses explicitly defined loss functions to bias the ultrasound field's optimization parameters to specific desired characteristics, such as axial resolution, transversal resolution, coverage, and focal region uniformity, while eliminating the uncertainty associated with virtual sources in time-reversal techniques. Numerical studies are conducted on four brain structures: the anterior insula, hippocampus, caudate nucleus, and amygdala. The findings are further validated in underwater experiments with a realistic 3D-printed skull phantom.

\end{abstract}


\maketitle
 \section{Introduction}

Transcranial focused ultrasound (tFUS) is a transformative technology that employs concentrated ultrasound energy to treat various neurological conditions. \cite{transcranial1,transcranial2}. Non-ionizing acoustic beams can penetrate through the skull and precisely target specific structures deep within the brain. The energy of the ultrasound beam elicits thermal and mechanical effects in brain tissue. Those effects are utilized for a variety of therapeutic treatments and procedures. At lower intensities, tFUS can accurately transmit energy to specific regions in the brain and induce local modulation of neural activity \cite{neuro1,neuro2,neuro3,neuro4}. Furthermore, tFUS has been demonstrated to locally disrupt the blood-brain barrier (BBB)  \cite{BBB1,BBB2,BBB3,BBB4}, allowing for improved delivery of therapeutic drugs. High-intensity ultrasound is also leveraged in tFUS to treat various diseases and disorders \cite{transcranial1}, for which ultrasound energy can be used to thermally ablate diseased tissue \cite{ablation1}, and high pressures can be used for histotripsy \cite{histo1,histo2,histo3}, in which targeted tissues are mechanically fractionated. 

Focusing the ultrasound beam in tFUS is challenging due to the highly stiff skull. The high acoustic impedance, irregular geometry, and nonuniform thickness of the skull bones strongly scatter and attenuate the propagating ultrasound beam. This phenomenon is known as skull-induced aberrations \cite{aberrations1,aberrations2}, and it causes severe defocusing and phase distortion of the beam, resulting in inefficient and out-of-target focusing. To address this issue, 3D-printed acoustic holographic lenses, known as acoustic holograms, have been recently proposed for tFUS and compensation of skull aberrations \cite{holotrans1,holotrans2,holotrans3,mouse,mouse2,skull,vortices}. Acoustic holograms act like phase plates, modulating the phase of the wavefront with unique thickness maps. The thickness map enables the reconstruction of complicated acoustic fields with a desired spatial distribution of pressure or phase using a single-element acoustic transducer. These lenses have been shown to modulate reflected and transmitted wavefronts \cite{sallamreflective,sallamreflective2,sallamNL,melde}, forming multi-focal spots or arbitrary complex patterns for applications such as contactless ultrasound energy transfer \cite{marjan}, particle manipulation \cite{melde,volumetric}, controlling cavitation zones \cite{cavitation}, and hyperthermia \cite{thermal}. These intricate modulation and manipulation capabilities make them ideal for tFUS applications. The acoustic lens compensates for skull aberrations and creates highly localized pressure fields that direct the ultrasound beam to specific brain areas within the skull. Acoustic holograms provide an order of magnitude higher resolution and acoustic pixel density than phased array transducers at a substantially lower cost and simpler fabrication procedure \cite{melde}. They have been tested successfully in bench-top setups using 3D-printed phantoms and ex-vivo human skulls \cite{skull,vortices}, as well as in-vivo mouse models \cite{mouse,mouse2} for neuromodulation and BBB opening procedures.

Time-reversal numerical simulations and direct error minimization techniques are currently employed to generate an appropriate thickness map that accounts for skull aberrations and directs the ultrasonic field to the target structure in the brain. In the numerical domain, monopole virtual sources are introduced at the target regions of the skull. The ultrasound is then transmitted from the simulated sources to the hologram plane outside of the skull, where the acoustic signal is recorded. The extracted signal is subsequently processed to determine the phase and amplitude distributions in the hologram plane. The phase and amplitude information are then used to construct the lens thickness map. Ultrasound propagation in strongly heterogeneous media is often carried out using full-wave numerical simulations, namely the pseudospectral time-domain method implemented in the open-source software k-Wave \cite{k-wave}. However, full-wave numerical simulations are computationally expensive, lasting up to 20 hours when using CPU parallel processing \cite{skull}. The long computation time also prevents the use of iterative optimization techniques, which could help improve focusing quality and allow the implementation of explicitly specified loss functions to bias the ultrasound field's optimization parameters to specific desired characteristics, such as axial resolution, transversal resolution, coverage, and focal region uniformity. Furthermore, iterative algorithms take into account the transducer's experimental amplitude output, improving the accuracy of the thickness map and eliminating the need for error diffusion schemes. Acoustic holography is a technique that allows for the reconstruction of 3-D acoustic fields by projecting measured or simulated 2-D data from a measurement plane to the remainder of the volume \cite{adaptive}. Because of its computing efficiency, holography is used in iterative optimization techniques such as the Gerchberg-Saxton (GS) and iterative angular spectrum algorithms to compute and design the thickness map of holographic lenses \cite{melde, volumetric}. In those techniques, the wavefront is iteratively propagated forward and backward between the hologram and target planes until a satisfying boundary condition is obtained. However, such techniques are intended for freely propagating media and optimizing planar pressure patterns. This paper introduces an analogous iterative optimization approach that takes into account strongly heterogeneous media and aberrating layers, including the effects of reflection, transmission, scattering, and attenuation. Furthermore, the proposed method optimizes for volumetric targets, such as directing sound to specific volumetric components in the brain. The modified mixed domain method (MMDM) \cite{strongly,weakly,msound} is used to simulate the forward and backward propagation inside the strongly heterogeneous media. In addition, a non-convex gradient descent optimization algorithm is used to find the optimal phase for targeting a specific volume. The proposed technique is substantially more efficient, allowing for the creation of a hologram in less than 50 minutes while also allowing for the explicit defining of loss functions. The proposed algorithm is evaluated numerically by considering four test cases that target different brain structures: the anterior insula, hippocampus, caudate nucleus, and amygdala. Each target resides at a different depth in the brain and has a distinct complex volume and geometric structure in 3D space. The holographic lenses obtained for each target are then 3D printed and tested in underwater trials and on a 3D-printed skull phantom. 

 \section{Methods}

  \subsection{Acoustic propagation}

 The MMDM \cite{msound,strongly,weakly} is implemented to execute computationally efficient forward and backward propagations through the strongly heterogeneous domain, which includes the skull and water. This methodology is based on the generalized Westervelt equation and can be thought of as an expanded form of the angular spectrum method. The acoustic field is calculated by marching in the spatial domain along the axis that is normal to the source plane. 

Using the generalized Westervelt equation as a starting point:
\begin{equation}
 \rho \nabla \cdot\left(\frac{1}{\rho} \nabla p\right)-\frac{1}{c^2} \frac{\partial^2 p}{\partial t^2}+\frac{\delta}{c^4} \frac{\partial^3 p}{\partial t^3}+\frac{\beta}{\rho c^4} \frac{\partial^2 p^2}{\partial t^2}=\gamma \frac{\partial p}{\partial t}   
\end{equation}

where $p$ is the acoustic pressure, $\rho$ is the ambient density, $c$ is the speed of sound, $\delta$ is the sound diffusivity and $\beta$ is the nonlinearity coefficient. The frequency-independent absorption term $\gamma \partial p / \partial t$ accounts for the absorption layer close to the computational boundaries.
Applying the normalized wavefield $f=p / \sqrt{\rho}$, assuming homogeneous density, and performing the Fourier transform with respect to $X$, $Y$, and $t$ yields:

\begin{equation}
\begin{aligned}
& \frac{\partial^{2}}{\partial Z^{2}} \tilde{P}+K^{2} \tilde{P}= \\
& F_{X Y}\left\{\left[-\frac{\omega^{2}}{c_{0}^{2}}\left(\frac{c_{0}^{2}}{c^{2}}-1\right)+\frac{i \delta \omega^{3}}{c^{4}}+i \omega \gamma\right] F_{t}(p)\right\} \\
& \quad+F_{X Y}\left(\frac{\beta \omega^{2}}{\rho c^{4}} F_{t}\left(p^{2}\right)\right)
\end{aligned}
\end{equation}

where $\tilde{P}$ is the Fourier transform (FT) of $p, F_{X Y}$ is the FT operator in $X$ - and $Y$-dimensions, $F_t$ is the FT operator in the time domain, $c_0$ is the background sound speed and $K^2=\omega^2 / c_0^2-k_X^2-k_Y^2$, $k_X$ and $k_Y$ are the wavenumbers in $X$ - and $Y$-dimensions, respectively. An implicit, one-way propagation can be derived from the 1-D Green's function in the form of an integral equation, such that
\begin{equation} \label{eq:2}
\tilde{P}(Z)=\tilde{P}(0) e^{i K Z}+\frac{e^{i K Z}}{2 i K} \int_0^Z e^{-i K Z^{\prime}} M\left(p\left(Z^{\prime}\right)\right) d Z^{\prime}
\end{equation}

where,
\begin{equation} 
\begin{aligned}
&M(p)=\\
& F_{X Y}\left\{\left[-\frac{\omega^{2}}{c_{0}^{2}}\left(\frac{c_{0}^{2}}{c^{2}}-1\right)+\frac{i \delta \omega^{3}}{c^{4}}+i \omega \gamma\right] F_{t}(p)\right\} \\
& +F_{X Y}\left(\frac{\beta \omega^{2}}{\rho c^{4}} F_{t}\left(p^{2}\right)\right)
\end{aligned}
\end{equation}

Equation \eqref{eq:2} is solved by using a Simpson-like rule. 

The pressure distribution on the source plane or any arbitrary plane is reconstructed using the backward projection. By changing $Z$ in \eqref{eq:2} to $-Z$, the backward projection is obtained as:
\begin{equation}
\tilde{P}(-Z)=\tilde{P}(0) e^{-i K Z}-\frac{e^{-i K Z}}{2 i K} \int_0^{-Z} e^{i K Z^{\prime}} M\left(p\left(-Z^{\prime}\right)\right) d Z^{\prime}
\end{equation}

The Kramers-Kronig dispersion relation is applied directly by replacing the speed of sound $c$ with $c_{p}$ and $c_{p}=\left(1 / \hat{c}+\alpha_{0} \tan (\pi y / 2) \omega^{y-1}\right)^{-1}$, where $\hat{c}$ is the sound speed at zero frequency, $y$ is the power-law exponent, $\alpha_{0}$ is the absorption in nepers per megahertz$^{-y}$ per meter and $\alpha_{o}=\alpha_{N P} \omega^{-y}$. Using this model, wave effects such as full-wave diffraction, attenuation, and dispersion are considered. However, this model is only accurate for weakly heterogeneous media \cite{weakly,strongly}. To have a more general model that could be applied to strongly heterogeneous media, phase correction and transmission compensation are introduced. 

Assuming 1D propagation through an inhomogeneous media with a varied speed
of sound distribution, the following phase correction term is added:

\begin{equation}
\begin{aligned}
\left(P_{z+\Delta z}\right)_{\text {corr}} = & \left(P_{z+\Delta z}\right) \\
 &+P_{z} e^{i K^{\prime} \Delta z}\left(1-e^{i\left(K-K^{\prime}\right) \Delta z} \frac{1+\frac{M_{z}}{4 i K} \Delta z}{1-\frac{M_{z+\Delta z}}{4 i K} \Delta z}\right)
\end{aligned}
\end{equation}

Where $P_{z}$ is the wave pressure at $z$, $(P_{z+\Delta z})_{\text {corr}}$ is the phase corrected pressure,   $K^{\prime}$ is the wave number, and $K^{\prime}=\omega / c_{2}$ with a frequency of $\omega$.

Moreover, the original formulation does not take into account the transmission coefficient caused by variations in sound speed. In order to handle the density and speed of sound heterogeneities, amplitude correction is employed. The compensation term:

\begin{equation}
\begin{aligned}
& T_corr(x, y, z) \\
& \quad=\frac{2 \rho(x, y, z+\Delta z) c(x, y, z+\Delta z)}{\rho(x, y, z) c(x, y, z)+\rho(x, y, z+\Delta z) c(x, y, z+\Delta z)}
\end{aligned}
\end{equation}

where $c(x, y, z)$ and $\rho(x, y, z)$ are the speed of sound and density at plane $z$, respectively, and $c(x, y, z+\Delta z)$ and $\rho(x, y, z+\Delta z)$ are the speed of sound and density at plane $z+\Delta z$, respectively.

Finally, reflections are further added to the formulation by using the following equation: 
\begin{equation}
    P_{\text {reflection }}=P_{\text {incident }}(T-1) \text {, }
\end{equation}{}

Here, $P_{\text {incident }}$ is the incident wave used for calculating the reflected wave. The second-order reflection field is formed by propagating the resulting $P_{\text {reflection }}$ in the forward direction. This process is repeated until the intended maximum order of reflection is achieved. This makes it possible to model scattering from multilayer interfaces and distributed scattering events. Superposing each solution yields the final wave field. 
 
  \subsection{Holograms computation}
  
Computer-generated holography (CGH) can be implemented to construct desired 3D pressure patterns by altering the wavefront of a coherent acoustic beam \cite{volumetric}. CGH determines the spatial phase and amplitude distribution of an incident wavefront, which, upon diffraction, forms the target volume. Computing the 2D phase mask that yields certain intensity patterns is generally a non-trivial and ill-posed task \cite{nonconvex}. Most three-dimensional intensity patterns are not physically viable. The wave field must diffract through space in a way that is consistent with Helmholtz's equations and maintains intensity conservation at each depth plane. Furthermore, resolution is limited due to the finite numerical aperture (NA). The phase-only hologram constraint also compounds these limitations further. As a result, any required 3D pressure pattern cannot be precisely reproduced. The optimal thickness map is determined by optimizing the output acoustic field under experimental limitations such as incident beam profile, source NA, and wave shaping modality. Following prior efforts in acoustics and optics, we use a gradient descent optimization approach in the present investigation \cite{nonconvex,volumetric}.

The defined loss function greatly influences the outcomes of 3D CGH, where there is no exact solution but numerous approximate solutions. Our algorithm searches for optimal holograms using custom cost functions that users can select to best suit a certain application. The optimization aims to obtain a phase distribution $\phi$ at the hologram plane that minimizes a specified error function $L[P_{3D}(\phi),T]$ between the generated acoustic field $P_{3D}(\phi)$ in the volumetric domain and the volumetric target pattern T, with the constraint that the pressure amplitude at the transducer plane must always equal to the experimental pressure $|p0|= A_0$. The numerical solution of this optimization problem necessitates calculating the gradient of the loss function with respect to the decision variables. By implementing Wirtinger subdifferential calculus and following the derivation in \cite{volumetric}, the gradient of the loss function $L$ is calculated as:

\begin{equation}
\begin{aligned}
 \nabla_{p} L=2 A^{*}\left(\left.\frac{A \boldsymbol{p}}{|A \boldsymbol{p}|} \circ \nabla_{x} \ell(\boldsymbol{x}, \boldsymbol{T})\right|_{x=|A \boldsymbol{p}|}\right)
\end{aligned}
\end{equation}

Here $A$ is a linear parameter representing the propagation of the acoustic field at the transducer plan to the rest of the volume, such that $A\boldsymbol{p}=P_{3D}$. Correspondingly, $A^{*}$ represents the inverse propagation from each plane in the volume to the transducer plane such that $A^{*}P_{3D}=\boldsymbol{p}$. Note that the propagation is implemented using the MMDM. Here $\circ$ denotes the element-wise product.
Assuming square loss

\begin{equation}
L(\boldsymbol{x}, \boldsymbol{T})=\frac{1}{2} \sum_{i=1}^{n_{x}} \sum_{j=1}^{n_{y}}\left(x_{i j}-T_{i j}\right)^{2}
\end{equation}

Substituting equation (10) into (9) yields:

\begin{equation}
\nabla_{p} L=2 A^{*}\left(A \boldsymbol{p}-\boldsymbol{T} \circ \frac{A \boldsymbol{p}}{|A \boldsymbol{p}|}\right)
\end{equation}

Thus the iterative gradient descent algorithm update rule becomes 

\begin{equation}
    p^{k+1}=\mathcal{P}_{U}\left[A^{*}\left(\boldsymbol{T} \circ \frac{A \boldsymbol{p}^{k}}{\left|A \boldsymbol{p}^{k}\right|}\right)\right]
\end{equation}

where $\mathcal{P}_{U}(\boldsymbol{p}):=\boldsymbol{u} \circ \boldsymbol{p} /|\boldsymbol{p}|$ is the Euclidean projection onto the constrain set $U$.

This algorithm is analogous to the GS algorithm and follows similar steps. However, with the aid of the MMDM, this algorithm correctly takes into account the heterogeneity of the media and uses adaptive step sizes with the aid of the Armijo line-search parameters $\beta$ and $\sigma$ as shown in Algorithm 1 below following \cite{volumetric}. In words, the algorithm's steps are as follows: (1) propagate the pressure from the transducer plane to the rest of the volume, (2) calculate the gradient of the loss function and back propagate each plane in the volume back to the hologram plane and obtain the new phase distribution at the transducer plane, (3) impose the new phase distribution and the initial transducer amplitude at the transducer plane, and (4) repeat steps 1-3 until the desired convergence criterion is met.  

The algorithm output provides a phase map for the holographic element, which we translate it into a surface map for fabrication. The resulting thickness map is then calculated as\cite{melde}: The holographic plate starts with an initial thickness $T_{0}$. Removing material at pixel position $(X, Y)$ causes a relative phase change

$$
\Delta \phi(X, Y)=\left(k_{\mathrm{m}}-k_{\mathrm{h}}\right) \Delta Th(X, X)
$$

Where $Th(X, Y)=Th_{0}-\Delta Th(X, Y)$ is the thickness of the pixel positioned at coordinates $(X, Y)$ in the hologram plane and $k_{\mathrm{h}}, k_{\mathrm{m}}$ are the wave numbers in the hologram plate and its surrounding medium, respectively.

\begin{algorithm}
\begin{algorithmic}
\State Result: Phase pattern $\phi = \arg(p)$
\State Initial field configuration $p = p_0$;
\State Initial stepsize $\tau = 1.0$;
\State Armijo line-search parameters $\beta, \sigma$;
\While{not converged}
    \State $\partial p=2 A^{*}\left(\left.\frac{A p}{\left|A p\right|} \circ \nabla_{x} \ell(x, T)\right|_{x=\left|A p\right|}\right)$
    \State $p_{\text{new}}=\mathcal{P}_{U}(p-\tau \partial p)$
    
    \While{$\ell\left(p_{\text{new}}\right)>\ell(p)-\sigma \frac{\tau}{2}\|\partial p\|^{2}$}
        \State $\tau=\beta \tau$
        \State $p_{\text{new}}=\mathcal{P}_{U}(p-\tau \partial p)$
    \EndWhile
    \State $p=p_{\text{new}}$
\EndWhile
\end{algorithmic}
\caption{Gradient descent with line search for computer-generated volumetric holography}
\end{algorithm}

\section{Numerical results}

\begin{figure*}
    \centering
    \includegraphics[width=1\linewidth]{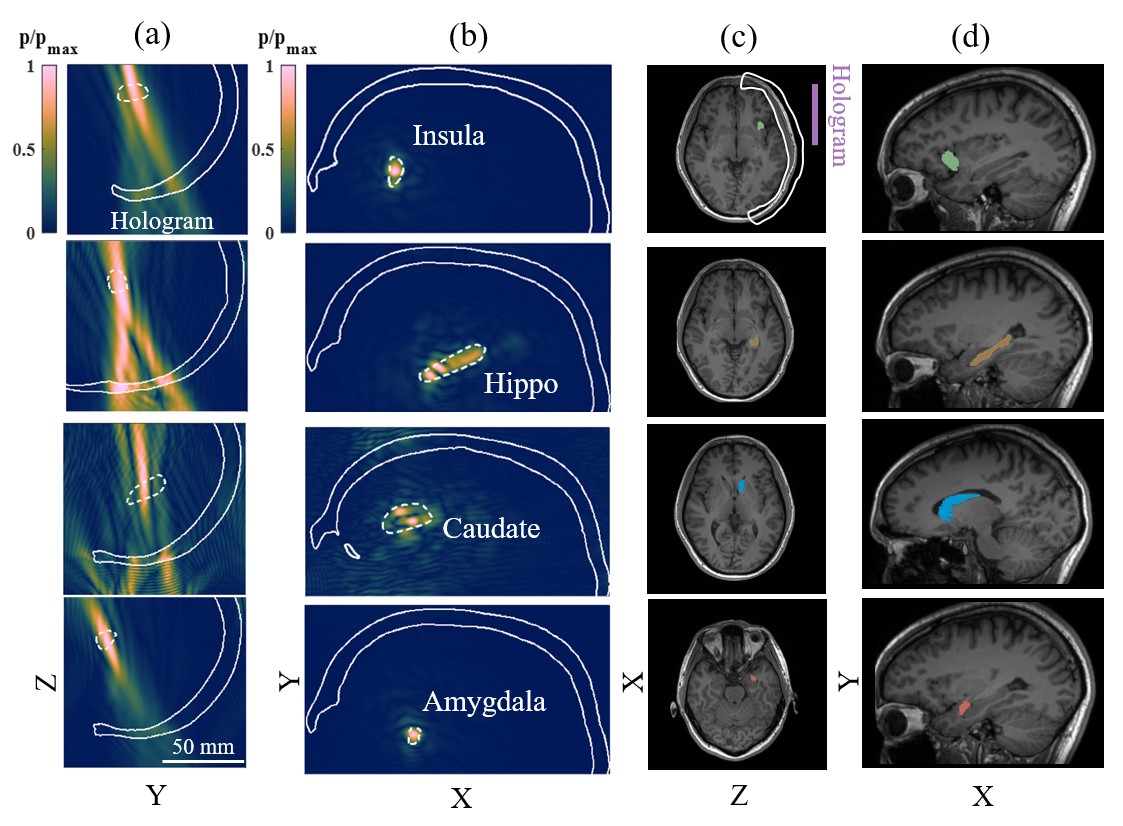}
    \caption{Summary of the numerical findings for targeting the desired volumes within the skull phantom. Columns (a) and (b) depict axial and transverse sections of the normalized pressure field, respectively. The dashed line represents the desired volume, whereas the solid white line represents the skull phantom. Columns (c) and (d) depict the axial and transverse sections of the MRI scans, respectively. The target volumes' geometries are emphasized in distinctive colors.}
    \label{fig2}
\end{figure*}

In this section, we numerically generate four acoustic holograms using the aforementioned optimization techniques to target four distinct human brain areas via the temporal window. A flat single-element transducer combined with an acoustic hologram can correct for skull aberrations and create a focal point tailored to a certain target location and volume. The geometry of the skull is derived from a computed tomography (CT) scan and processed to a resolution of the numerical grid. The skull geometry is sliced along a plane parallel to the sagittal plane to match the 3D-printed skull in the experimental setting. The skull section is integrated into the numerical simulations with a resolution of $0.5$ mm, which corresponds to $\lambda/6.67$ for the transducer drive frequency in water. The numerical grid's resolution is uniform in all directions ($dX=dY=dZ$). The geometry and placement of the therapeutic targets chosen for this investigation have been retrieved from the Magnetic resonance imaging (MRI) scans and fitted and mapped into their proper locations in the cranial cavity. The flat circular transducer and hologram are situated at the temporal bone window, at a distance of  $15 \mathrm{~mm}$ between the hologram's surface and the skull's surface. The transducer's output is experimentally evaluated in a water tank, with hydrophone measurements taken at an excitation frequency of $444$ kHz. The numerical algorithm is constrained by the transducer's measured experimental amplitude, and the thickness map is designed using the initial experimental phase, as detailed in Algorithm 1. In all simulations, the average amplitude at the transducer's surface is normalized to $1\mathrm{Pa}$. Water is assumed everywhere else in the numerical domain since it is a good approximation of soft tissue and to match the water-submerged experiments. The domain size is $384X212X199$ grid points in the X, Y, and Z directions, respectively. The Armijo line-search parameters in Algorithm 1 are set to $\beta=0.9$ and $\sigma=0.002$ with an initial step size of $\tau=1$. The target volume $T$ in Algorithm 1 is a binary volumetric matrix with nonzero values at the targeted brain structure. The numerical simulations of acoustic propagation are performed using the MMDM and implemented in the open-source toolbox mSOUND on MATLAB R2019b \cite{msound}. A computational node equipped with AMD EPYC 7702 2.5 GHz (total of 18 cores) and 36 GB of RAM is used for all simulations. The backward propagation step in Algorithm 1 is the most computationally time-consuming. In Algorithm 1, the backward propagation stage consumes the most processing time. The pressure at each plane in the entire volume must be propagated back to the transducer plane. Because each propagation is independent, a parallel backpropagation technique was created using MATLAB's $parfor$ function, which considerably accelerated the procedure. In the current study, each iteration lasted an average of 380 seconds, and each simulation produced satisfactory results after seven iterations on average. The average overall simulation time was 44 minutes, which is much faster than previously reported time-reversal techniques.

To assess the proposed algorithm's performance, we investigate four separate test cases involving various targets inside the brain. We created four distinct holograms that focus the ultrasound field on the anterior insula, hippocampus, caudate nucleus, and amygdala. Each of these targets has distinct properties in terms of shape, volume, orientation, and spatial placement within the brain structure. This diversity allows for a more comprehensive and general assessment of the holograms' performance capabilities. We shall look at the axial and transversal distribution of the ultrasonic field within the brain. In addition, we look into the sonication volume and peak focal pressure gain inside and outside the target regions. Sonication volumes are calculated as the region with a pressure gain of at least $-3 \mathrm{~dB}$ with respect to the peak pressure in the brain for each test instance. Normalized peak pressure is calculated as the pressure gain relative to the average pressure at the transducer surface. The results of the numerical simulations for all the test cases are summarised in Fig. \ref{fig2} and quantitatively in Table \ref{tab:my_table}.

\subsection{Anterior Insula}
The anterior insula is a brain area that processes taste, emotions, and interoceptive awareness, including hunger and pain. It also plays roles in autonomic functioning and sensory processing, including auditory and vestibular functions, and is essential for empathy and social cognition \cite{insula1,insula2}. In the present simulation, the insula's volume is $0.875$ cm$^3$, and it is positioned at a depth of $8.32$ cm from the hologram's surface along the axial direction, making it the deepest target among all the test cases. The dimensions of the insula extend to lengths of $0.9$ cm, $1.9$ cm, and $1$ cm along the X, Y, and Z axes, respectively. The normalized cross-section of the maximum pressure field in the YZ and XY planes are shown in the first row of Fig. \ref{fig2} (a) and (b), respectively. Each cross-section intersects the centroid of the target volume. The cartesian axes are aligned with the axes of the source. The segmented MR image with the highlighted geometry of the anterior insula is illustrated in \ref{fig2} (c) and (d). As illustrated in the axial and transversal cross-sections, the pressure is concentrated on the Insula volume, which is marked by the white dashed line. Good coverage is obtained, with the main focal encompassing the majority of the volume. However, while the energy is focused on the Insula, significant acoustic pressure occurs both before and after the focal zone. The maximum normalized pressures inside and outside the target volume are close, at a value of $4.55$ and $4.67$, respectively. A total of $0.45$ $\mathrm{cm}^{3}$ of the Insula was sonicated, representing $51.4 \%$ of the total volume. While $2.25$ $\mathrm{cm}^{3}$ was sonicated outside the target. It is also crucial to note the low-pressure amplitudes seen in the skull (which is outlined by solid white lines). This is crucial since it reduces the possibility of skull heating effects, which improves the therapeutic technique's safety profile.

\subsection{Hippocampus}
The hippocampus, located in the temporal lobe of the brain, is responsible for memory formation and storage, spatial navigation, and stress and mood regulation. It also affects the learning process. Dysfunction in this area can lead to memory issues, spatial disorientation, and emotional problems \cite{hippocampus}. The volume of the segmented Hippocampus is $3.85$ cm$^3$, and it is located at a depth of $7.55$ cm from the hologram's surface along the axial direction. The dimensions of the Hippocampus extend to lengths of $4.45$ cm, $2.55$ cm, and $1.55$ cm along the X, Y, and Z axes, respectively. The normalized cross-section of the maximum pressure field in the YZ and XY planes are shown in the second row of Fig. \ref{fig2} (a) and (b), respectively. MRI scans, presented in  Fig. \ref{fig2} (c) and (d), show that the hippocampus extends more prominently in the transversal plane along the X-axis. At 4.45 cm, the hippocampus is the largest in terms of length in the X direction among all the target volumes investigated. The axial and transverse cross-sections clearly show the pressure focus within the hippocampus volume. The transverse cross-section shows that pressure is effectively contained inside the intended area. However, axially, substantial pressure exists outside the target zone. The maximum normalized pressure inside and outside the target volume are close, at a value of 2.97 and 3.15, respectively. Using the hologram $1.61$ $\mathrm{cm}^{3}$ or $41.8 \%$ of the Hippocampus is sonicated. However, a significant portion of the sonicated volume, amounting to $8.1$ $\mathrm{cm}^{3}$, occurs outside the targeted region, with $83.4 \%$ of the total sonicated volume falling outside the intended area. Furthermore, high-pressure amplitudes in the skull increase the risk of causing unwanted skull heating effects when performing therapeutic operations.

\subsection{Caudate Nucleus}
The caudate nucleus in the brain's basal ganglia is crucial for motor control, procedural learning, reward processing, decision-making, and emotional processing. Its dysfunction is associated with various neurological and psychiatric disorders, including Parkinson's and Huntington's diseases \cite{Caudate}. The volume of the segmented caudate nucleus is $4.73$ cm$^3$, and it is positioned at a depth of $6.2$ cm from the hologram's surface along the axial direction, making it the largest target and the closet to the hologram surface among all the targets considered in the present investigation. The dimensions of the caudate extend to lengths of $4.1$ cm, $2.3$ cm, and $1.75$ cm along the X, Y, and Z axes, respectively. The normalized cross-section of the maximum pressure field in the YZ and XY planes are shown in the third row of Fig. \ref{fig2} (a) and (b), respectively. MRI scans, as presented in Fig. \ref{fig2} (c) and (d), show that the caudate also extends more prominently in the transversal plane along the X-axis. The axial and transverse cross-sections show the pressure concentration inside the caudate's volume, however there is less effective coverage of the target area than in earlier tests. This is emphasized by the relatively smaller sonicated volume of the caudate, amounting to $1.2$ $\mathrm{cm}^{3}$ or $25.4 \%$ of its total volume. This level of sonication is notably the lowest among all the considered test cases. The sonicated volume outside of the target volume is $8.81$ $\mathrm{cm}^{3}$ which is $86.3\%$ of the total sonicated volume. The maximum normalized pressure inside and outside the target volume are close, at a value of $3.32$ and $3.38$, respectively.

\subsection{Amygdala}

The amygdala, a tiny region in the brain, is critical in emotional processing, notably fear and aggression. It is essential for developing emotional memories and influencing decision-making. The amygdala is also involved in stress and anxiety reactions, as well as social interactions through the interpretation of social cues. Amygdala dysfunction can cause emotional disorders, memory problems, and difficulties with social behavior\cite{Amygdala}. The volume of the segmented Amygdala is $0.491$ cm$^3$, and it is positioned at a depth of $7.49$ cm from the hologram's surface along the axial direction. The dimensions of the amygdala extend to lengths of $0.8$ cm, $1.2$ cm, and $1.15$ cm along the X, Y, and Z axes, respectively. The amygdala is the smallest and most contained target volume among all the test cases, as highlighted by the MRI scans in Fig. \ref{fig2} (c) and (d). The normalized cross-section of the maximum pressure field in the YZ and XY planes are shown in the fourth row of Fig. \ref{fig2} (a) and (b), respectively. The axial and transversal cross-sections demonstrate adequate coverage and containment of the focus zone in the target region. This is highlighted by the high sonication rate within the amygdala with a total sonication volume of $0.32$ $\mathrm{cm}^{3}$ or $65.3\%$, which is the highest sonication percentage among all the test cases. The sonicated volume outside of the target volume is $1.93$ $\mathrm{cm}^{3}$, which represents $85.8\%$ of the total sonicated volume. The maximum normalized pressure inside and outside the target volume is the same, at 5.49, which is the largest gain achieved among all test cases. Low-pressure amplitudes in the skull were also observed when targeting the amygdala.

\begin{table*}[htbp]
    \caption{Summary of the Numerical Results}
    \label{tab:my_table}
    \centering
    \begin{tabularx}{\textwidth}{*{13}{>{\centering\arraybackslash}X}} % 13 columns
        \toprule
          & Volume [$\mathrm{cm}^{3}$] & Total sonicated volume [$\mathrm{cm}^{3}$] & Sonication inside target [$\mathrm{cm}^{3}$] & Percent of target sonication [\%] & Gain inside target & Gain outside target \\
        \midrule
        Insula & 0.875 & 2.7 & 0.45 & 51.4 & 4.55 & 4.67 \\
        Hippocampus & 3.85 & 9.9 & 1.61 & 41.8 & 2.97 & 3.15 \\
        Caudate & 4.73 & 8.81 & 1.2 & 25.4 & 3.32 & 3.38 \\
        Amygdala & 0.491 & 2.22 & 0.321 & 65.3 & 5.49 & 5.49 \\
        \bottomrule
    \end{tabularx}
\end{table*}

\subsection{Discussion}

The pressure fields created by acoustic holograms are inherently constrained by the diffraction limit and must adhere to the Helmholtz equation and the principle of energy conservation. This implies that sound fields that do not conform to these fundamental principles are not physically viable. As a result, there will inevitably be differences between the theoretically desired target field and the actual field that is physically realized due to these inherent limitations and principles governing acoustic propagation. The sound field can be significantly improved and optimized by carefully considering key parameters such as the numerical aperture, operating frequency, and the location and orientation of both the transducer and the hologram in relation to the size, position, orientation, and geometry of the target volume. This rigorous calibration guarantees that the acoustic field is fine-tuned to meet the appropriate targeting characteristics, maximizing the treatment process's efficacy, precision, and safety.

The numerical results clearly show that acoustic holograms are more effective at restricting the focal region within the transversal portion than in the axial section, especially for bigger targets such as the hippocampus and caudate. The wider the target's extent and cross-sectional area in the transversal plane, the weaker the field's focus is, resulting in a lower gain and a focused zone that is more elongated and distributed in the axial direction, causing greater sonication outside the target volume. This impact can be minimized by increasing the numerical aperture (transducer size) or adjusting the operating frequency, albeit both of these modifications have limitations. Another technique is to orient the transducer so that its axis is parallel to the target's longest dimension, resulting in an elongated focal area that is substantially contained inside the target volume. However, this strategy has limitations; for example, targeting the hippocampus will necessitate passing through the parietal bone, which is thicker and more curved, which could lead to new complications and issues such as skull heating and mode convergence.

These constraints are inherent in acoustic holographic lenses, not the optimization algorithm itself. However, the optimization methodology provided here, with its explicitly defined loss function, provides a method for somewhat overcoming these limits. Different loss and weighting functions can be used to direct the optimization algorithm to prioritize specific ultrasonic parameters. For example, to promote axial resolution by giving axial errors more weight than transversal errors. This approach could provide better axial confinement at the expense of transversal containment. Other parameters, such as pressure uniformity and coverage, can also be optimized to fit specific therapeutic demands and objectives, resulting in a more specialized optimization process that improves the efficacy of ultrasonic applications. The loss functions can be customized further by designating regions with low pressure, directing the algorithm to avoid sonication in these areas. This adjustment provides more precise control over pressure distribution, protecting vital or sensitive areas from unwanted ultrasonic wave exposure. Furthermore, the proposed algorithm effectively eliminates the uncertainties inherent in time-reversal techniques, where the impact of the number, distribution, and locations of virtual sources on focusing parameters is unclear. Determining the optimal distribution of virtual sources for specific desired focusing parameters can be challenging due to the non-iterative nature of these techniques and the computational limitations associated with them \cite{numerical}.

\section{Experimental verification}

\subsection{Experimental Setup}
\begin{figure}
    \centering
    \includegraphics[width=1\linewidth]{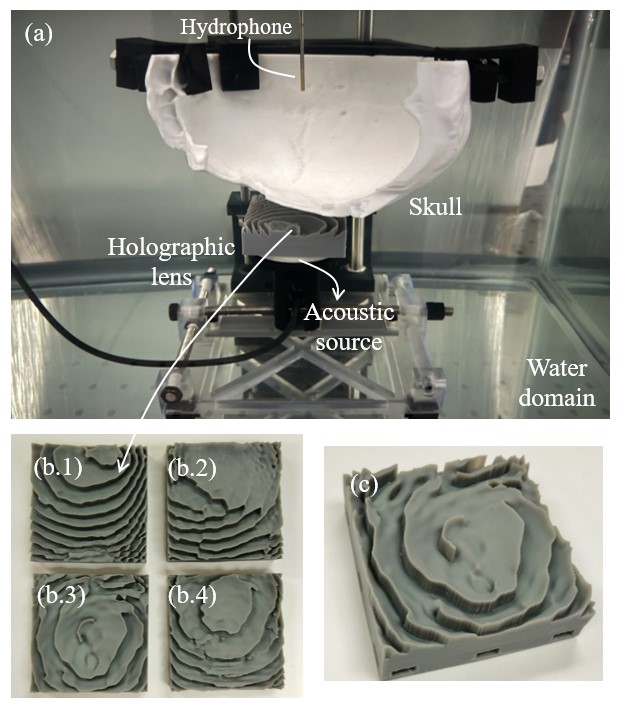}
    \caption{(a) shows the experimental setup, including the acoustic source, 3D-printed acoustic hologram, 3D-printed skull phantom, and the hydrophone. The 3D-printed holograms for targetting the insula (b.1), hippocampus (b.2), caudate (b.3), and the amygdala (b.4). (c) closeup of the 3D-printed hologram.}
    \label{fig1}
\end{figure}

The experimental setup is submerged in a water tank, as shown in Fig. \ref{fig1}. The walls of the tank are partially covered with Aptflex F28 acoustic absorbing sheets manufactured by Precision Acoustics, Ltd. The water tank is filled with degassed and deionized water in a temperature-controlled environment. A custom-made holder is fabricated to hold the skull, the ultrasound transducer, and the holographic lens. The position of the transducer can be moved manually relative to the skull in the X,Y, and Z directions. The transducer is custom-made and composed of a single circular piezoelectric disc with a diameter of $60$ mm and an operating frequency of $444$ kHz. The transducer is excited with a sinusoidal pulse signal (20 cycles) by a signal generator (Keysight 33500B) and an amplifier (Electronics and Innovation A075). The pressure field on a specified target plane is mapped with the ONDA HNR–0500 needle hydrophone attached to a custom-built in-house 3D positioning system. The measured signal from the hydrophone is sampled and digitized by a Tektronix TBS2104 oscilloscope and averaged 64 times to increase the signal-to-noise ratio. To eliminate the effect of any DC bias and higher frequency noises, the signal is post-processed using MATLAB by performing an FFT on the time series. A MATLAB program is used to control the positioning system and triggering of the function generator. The obtained voltage amplitude is then converted into acoustic pressure according to the calibration factor provided by the manufacturer (-250.13 dB). The geometry of the skull is obtained using a segmented CT scan. A section of the skull is then 3D-printed with Polycarbonate (PC) FDM  by STRATASYS (F270, Eden Prairie, MN) with a smooth hand finish. The acoustic properties of PC have been previously investigated underwater \cite{thermoplastics} with $c=2380$ m/s, $\rho=1153$ kg/m$^3$, and $\alpha=12.23$ dB/cm at $1.1$ MHz. A Formlabs (Formlabs, Somerville, MA, US) Form 3 grey resin was used for printing the acoustic lenses with a lateral and axial resolution of 25 microns. The acoustic properties of the resin were previously reported in an underwater experimental study \cite{photopolymers} with $c=2591$ m/s, $\rho=1178$ kg/m$^3$, and $\alpha=2.922$ dB/cm at $1$ MHz.  

\begin{figure*}
    \centering
    \includegraphics[width=0.8\linewidth]{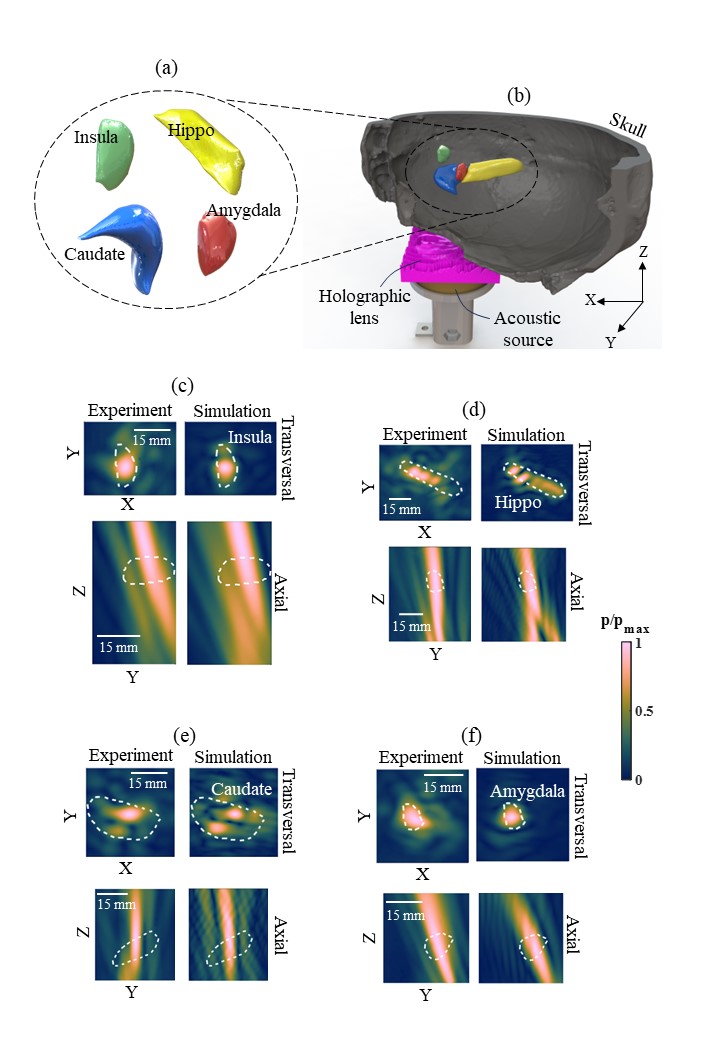}
    \caption{(a) The geometry of each target evaluated in the current study. (b) Schematic of the experimental setup, with the targets positioned in their proper placements within the cranial cavity of the skull phantom. (c), (d), (e), and (f) show the experimental and numerical results for the test cases of the insula, hippocampus, caudate, and amygdala.}
    \label{fig3}
\end{figure*}

\subsection{Experimental results}

Four holograms are designed based on the numerical results to validate the proposed algorithm. Underwater pressure scans in both the transversal and axial directions are carried out for each target using a hydrophone attached to a  positioning system. Due to the hydrophone's limited mobility range within the skull, only smaller portions surrounding the target location are scanned. The experimental measurements are then compared to numerical simulations.
Figure \ref{fig3} (a) and (b), respectively, show the geometry of the four target volumes and their corresponding location inside the 3D-printed skull phantom. Figure \ref{fig3} (c), (d), (e), and (f) respectively compare the experimental measurements and their corresponding numerical results for the insula, hippocampus, caudate, and amygdala. Good overall agreement is observed between the experimental measurements and the numerical results. The maximum experimental pressures measured in the transversal plane were $85.2$ kPa for the insula, $98.5$ kPa for the Hippocampus, $127$ kPa for the Caudate, and $98.2$ kPa for the Amygdala with an excitation input of $200$ V$_{pp}$. 

For a quantitative comparison between experimental and numerical results, we calculated the focusing efficiency \cite{sallamreflective} in the transversal section. By comparing the acoustic power within the focal region to the total acoustic power in the entire plane. The calculations assume planar wave propagation. The focusing efficiency is given as:
\begin{equation}
    \eta = \frac{\Pi_{focal}}{\Pi_{total}}=\frac{\sum P_{focal}^2(X,Y,Z_0)}{\sum P^2(X,Y,Z_0)}
\end{equation}
Where $P_{focal} (X,Y,Z_0)$ and $\Pi_{focal}$ are the acoustic pressure distribution and the acoustic power in the focal regions, respectively. $\Pi_{total}$ is the total acoustic power in the transversal plane.  
Examining Fig. \ref{fig3} (c), which presents insula results, we notice more pressure leakage outside of the focal region in the experimental data. This is evident from the disparity in focusing efficiency, with the simulations calculated at $76.5\%$ compared to $46.1\%$ in the experimental measurements. A similar pattern is observed in other test cases: the hippocampus has a numerical efficiency of $84.2\%$ compared to an experimental efficiency of $49.2\%$, and the amygdala with a numerical efficiency of $70.1\%$ against an experimental efficiency of $41.6\%$. Interestingly, the caudate is the only test case where the experimental results outperform the simulations, with an efficiency of $64.2\%$ versus $57.1\%$, potentially due to a more pronounced top lobe in the experimental data as shown in Fig. \ref{fig3}(e). The differences between experimental and numerical results could be attributable to factors such as measurement mistakes, alignment challenges, variances in material qualities, and flaws in the produced skull phantom and hologram.

\section{Conclusions}

We introduced an efficient volumetric holographic technique for creating 3D-printed acoustic holograms specifically for low-intensity non-thermal tFUS applications. The fundamental goal of these acoustic lenses is to generate precise FUS fields within the brain while successfully compensating for aberrations induced by the skull's structure. The proposed optimization strategy uses the modified mixed domain method to perform efficient ultrasonic propagation in strongly heterogeneous media, which is then incorporated into a gradient descent iterative volumetric holography optimization algorithm. The holographic approach outperforms the previously reported time-reversal strategies used in full-wave time-domain numerical simulations. Using the proposed approach, a hologram can be created in less than 50 minutes. Furthermore, due to the more efficient propagation, iterative optimization techniques can be used with explicitly assigned loss functions. The optimization technique can be fine-tuned to prioritize individual ultrasound characteristics by using different loss and weighting functions. For example, increasing axial resolution entails giving greater weight to axial errors than transversal ones, which improves axial containment but may reduce transversal containment. Other factors like pressure uniformity and coverage can also be adjusted to suit particular therapeutic requirements, enabling a more customized optimization process. The loss functions can also be specifically tailored to designate regions where minimal pressure is crucial, thereby directing the algorithm to prevent sonication in those sensitive areas, ensuring a more precise and effective ultrasound treatment. We applied the proposed algorithm to target four brain structures: the anterior insula, hippocampus, caudate nucleus, and amygdala. Verification was done through underwater experiments using realistic 3D-printed skull phantoms. Good agreement was observed between the numerical results and the measured experimental data. Our results demonstrate that the acoustic holograms achieved effective coverage and confinement in the transversal plane. However, in the axial direction, sonication leakage varied depending on the target geometry, orientation, and volume, with lower axial containment observed in larger and transversely elongated targets like the hippocampus and caudate. The proposed approach enables more rapid computation of acoustic holograms, enabling quicker administration of therapeutic procedures and allowing for more flexibility in tailoring ultrasound fields for specific therapeutic requirements.

\section*{ACKNOWLEDGEMENT}
\vspace{-15pt}
This work was supported by the U.S. National Science Foundation (NSF) under CAREER grant, Award No. CMMI $2143788$, which is gratefully acknowledged.
\vspace{15pt}
\section*{DATA AVAILABILITY}
\vspace{-15pt}
The data that support the findings of this study are available from the corresponding author upon reasonable request.
\vspace{-15pt}	
\section*{REFERENCES}
\vspace{-15pt}
\bibliography{MAIN}% Produces the bibliography via BibTeX.

\providecommand{\noopsort}[1]{}\providecommand{\singleletter}[1]{#1}%
\begin{thebibliography}{45}%
\makeatletter
\providecommand \@ifxundefined [1]{%
 \@ifx{#1\undefined}
}%
\providecommand \@ifnum [1]{%
 \ifnum #1\expandafter \@firstoftwo
 \else \expandafter \@secondoftwo
 \fi
}%
\providecommand \@ifx [1]{%
 \ifx #1\expandafter \@firstoftwo
 \else \expandafter \@secondoftwo
 \fi
}%
\providecommand \natexlab [1]{#1}%
\providecommand \enquote  [1]{``#1''}%
\providecommand \bibnamefont  [1]{#1}%
\providecommand \bibfnamefont [1]{#1}%
\providecommand \citenamefont [1]{#1}%
\providecommand \href@noop [0]{\@secondoftwo}%
\providecommand \href [0]{\begingroup \@sanitize@url \@href}%
\providecommand \@href[1]{\@@startlink{#1}\@@href}%
\providecommand \@@href[1]{\endgroup#1\@@endlink}%
\providecommand \@sanitize@url [0]{\catcode `\\12\catcode `\$12\catcode `\&12\catcode `\#12\catcode `\^12\catcode `\_12\catcode `\%12\relax}%
\providecommand \@@startlink[1]{}%
\providecommand \@@endlink[0]{}%
\providecommand \url  [0]{\begingroup\@sanitize@url \@url }%
\providecommand \@url [1]{\endgroup\@href {#1}{\urlprefix }}%
\providecommand \urlprefix  [0]{URL }%
\providecommand \Eprint [0]{\href }%
\providecommand \doibase [0]{http://dx.doi.org/}%
\providecommand \selectlanguage [0]{\@gobble}%
\providecommand \bibinfo  [0]{\@secondoftwo}%
\providecommand \bibfield  [0]{\@secondoftwo}%
\providecommand \translation [1]{[#1]}%
\providecommand \BibitemOpen [0]{}%
\providecommand \bibitemStop [0]{}%
\providecommand \bibitemNoStop [0]{.\EOS\space}%
\providecommand \EOS [0]{\spacefactor3000\relax}%
\providecommand \BibitemShut  [1]{\csname bibitem#1\endcsname}%
\let\auto@bib@innerbib\@empty
%</preamble>
\bibitem [{\citenamefont {Meng}, \citenamefont {Hynynen},\ and\ \citenamefont {Lipsman}(2021)}]{transcranial1}%
  \BibitemOpen
  \bibfield  {author} {\bibinfo {author} {\bibfnamefont {Y.}~\bibnamefont {Meng}}, \bibinfo {author} {\bibfnamefont {K.}~\bibnamefont {Hynynen}}, \ and\ \bibinfo {author} {\bibfnamefont {N.}~\bibnamefont {Lipsman}},\ }\href@noop {} {\bibfield  {journal} {\bibinfo  {journal} {Nature Reviews Neurology}\ }\textbf {\bibinfo {volume} {17}},\ \bibinfo {pages} {7} (\bibinfo {year} {2021})}\BibitemShut {NoStop}%
\bibitem [{\citenamefont {Darmani}\ \emph {et~al.}(2022)\citenamefont {Darmani}, \citenamefont {Bergmann}, \citenamefont {Pauly}, \citenamefont {Caskey}, \citenamefont {De~Lecea}, \citenamefont {Fomenko}, \citenamefont {Fouragnan}, \citenamefont {Legon}, \citenamefont {Murphy}, \citenamefont {Nandi} \emph {et~al.}}]{transcranial2}%
  \BibitemOpen
  \bibfield  {author} {\bibinfo {author} {\bibfnamefont {G.}~\bibnamefont {Darmani}}, \bibinfo {author} {\bibfnamefont {T.}~\bibnamefont {Bergmann}}, \bibinfo {author} {\bibfnamefont {K.~B.}\ \bibnamefont {Pauly}}, \bibinfo {author} {\bibfnamefont {C.}~\bibnamefont {Caskey}}, \bibinfo {author} {\bibfnamefont {L.}~\bibnamefont {De~Lecea}}, \bibinfo {author} {\bibfnamefont {A.}~\bibnamefont {Fomenko}}, \bibinfo {author} {\bibfnamefont {E.}~\bibnamefont {Fouragnan}}, \bibinfo {author} {\bibfnamefont {W.}~\bibnamefont {Legon}}, \bibinfo {author} {\bibfnamefont {K.}~\bibnamefont {Murphy}}, \bibinfo {author} {\bibfnamefont {T.}~\bibnamefont {Nandi}},  \emph {et~al.},\ }\href@noop {} {\bibfield  {journal} {\bibinfo  {journal} {Clinical Neurophysiology}\ }\textbf {\bibinfo {volume} {135}},\ \bibinfo {pages} {51} (\bibinfo {year} {2022})}\BibitemShut {NoStop}%
\bibitem [{\citenamefont {Legon}\ \emph {et~al.}(2014)\citenamefont {Legon}, \citenamefont {Sato}, \citenamefont {Opitz}, \citenamefont {Mueller}, \citenamefont {Barbour}, \citenamefont {Williams},\ and\ \citenamefont {Tyler}}]{neuro1}%
  \BibitemOpen
  \bibfield  {author} {\bibinfo {author} {\bibfnamefont {W.}~\bibnamefont {Legon}}, \bibinfo {author} {\bibfnamefont {T.~F.}\ \bibnamefont {Sato}}, \bibinfo {author} {\bibfnamefont {A.}~\bibnamefont {Opitz}}, \bibinfo {author} {\bibfnamefont {J.}~\bibnamefont {Mueller}}, \bibinfo {author} {\bibfnamefont {A.}~\bibnamefont {Barbour}}, \bibinfo {author} {\bibfnamefont {A.}~\bibnamefont {Williams}}, \ and\ \bibinfo {author} {\bibfnamefont {W.~J.}\ \bibnamefont {Tyler}},\ }\href@noop {} {\bibfield  {journal} {\bibinfo  {journal} {Nature neuroscience}\ }\textbf {\bibinfo {volume} {17}},\ \bibinfo {pages} {322} (\bibinfo {year} {2014})}\BibitemShut {NoStop}%
\bibitem [{\citenamefont {Lee}\ \emph {et~al.}(2016)\citenamefont {Lee}, \citenamefont {Kim}, \citenamefont {Jung}, \citenamefont {Chung}, \citenamefont {Song}, \citenamefont {Lee},\ and\ \citenamefont {Yoo}}]{neuro2}%
  \BibitemOpen
  \bibfield  {author} {\bibinfo {author} {\bibfnamefont {W.}~\bibnamefont {Lee}}, \bibinfo {author} {\bibfnamefont {H.-C.}\ \bibnamefont {Kim}}, \bibinfo {author} {\bibfnamefont {Y.}~\bibnamefont {Jung}}, \bibinfo {author} {\bibfnamefont {Y.~A.}\ \bibnamefont {Chung}}, \bibinfo {author} {\bibfnamefont {I.-U.}\ \bibnamefont {Song}}, \bibinfo {author} {\bibfnamefont {J.-H.}\ \bibnamefont {Lee}}, \ and\ \bibinfo {author} {\bibfnamefont {S.-S.}\ \bibnamefont {Yoo}},\ }\href@noop {} {\bibfield  {journal} {\bibinfo  {journal} {Scientific reports}\ }\textbf {\bibinfo {volume} {6}},\ \bibinfo {pages} {34026} (\bibinfo {year} {2016})}\BibitemShut {NoStop}%
\bibitem [{\citenamefont {Mueller}\ \emph {et~al.}(2014)\citenamefont {Mueller}, \citenamefont {Legon}, \citenamefont {Opitz}, \citenamefont {Sato},\ and\ \citenamefont {Tyler}}]{neuro3}%
  \BibitemOpen
  \bibfield  {author} {\bibinfo {author} {\bibfnamefont {J.}~\bibnamefont {Mueller}}, \bibinfo {author} {\bibfnamefont {W.}~\bibnamefont {Legon}}, \bibinfo {author} {\bibfnamefont {A.}~\bibnamefont {Opitz}}, \bibinfo {author} {\bibfnamefont {T.~F.}\ \bibnamefont {Sato}}, \ and\ \bibinfo {author} {\bibfnamefont {W.~J.}\ \bibnamefont {Tyler}},\ }\href@noop {} {\bibfield  {journal} {\bibinfo  {journal} {Brain stimulation}\ }\textbf {\bibinfo {volume} {7}},\ \bibinfo {pages} {900} (\bibinfo {year} {2014})}\BibitemShut {NoStop}%
\bibitem [{\citenamefont {Legon}\ \emph {et~al.}(2018{\natexlab{a}})\citenamefont {Legon}, \citenamefont {Ai}, \citenamefont {Bansal},\ and\ \citenamefont {Mueller}}]{neuro4}%
  \BibitemOpen
  \bibfield  {author} {\bibinfo {author} {\bibfnamefont {W.}~\bibnamefont {Legon}}, \bibinfo {author} {\bibfnamefont {L.}~\bibnamefont {Ai}}, \bibinfo {author} {\bibfnamefont {P.}~\bibnamefont {Bansal}}, \ and\ \bibinfo {author} {\bibfnamefont {J.~K.}\ \bibnamefont {Mueller}},\ }\href@noop {} {\bibfield  {journal} {\bibinfo  {journal} {Human brain mapping}\ }\textbf {\bibinfo {volume} {39}},\ \bibinfo {pages} {1995} (\bibinfo {year} {2018}{\natexlab{a}})}\BibitemShut {NoStop}%
\bibitem [{\citenamefont {Deffieux}\ and\ \citenamefont {Konofagou}(2010)}]{BBB1}%
  \BibitemOpen
  \bibfield  {author} {\bibinfo {author} {\bibfnamefont {T.}~\bibnamefont {Deffieux}}\ and\ \bibinfo {author} {\bibfnamefont {E.~E.}\ \bibnamefont {Konofagou}},\ }\href@noop {} {\bibfield  {journal} {\bibinfo  {journal} {IEEE transactions on ultrasonics, ferroelectrics, and frequency control}\ }\textbf {\bibinfo {volume} {57}},\ \bibinfo {pages} {2637} (\bibinfo {year} {2010})}\BibitemShut {NoStop}%
\bibitem [{\citenamefont {Legon}\ \emph {et~al.}(2018{\natexlab{b}})\citenamefont {Legon}, \citenamefont {Ai}, \citenamefont {Bansal},\ and\ \citenamefont {Mueller}}]{BBB2}%
  \BibitemOpen
  \bibfield  {author} {\bibinfo {author} {\bibfnamefont {W.}~\bibnamefont {Legon}}, \bibinfo {author} {\bibfnamefont {L.}~\bibnamefont {Ai}}, \bibinfo {author} {\bibfnamefont {P.}~\bibnamefont {Bansal}}, \ and\ \bibinfo {author} {\bibfnamefont {J.~K.}\ \bibnamefont {Mueller}},\ }\href@noop {} {\bibfield  {journal} {\bibinfo  {journal} {Human brain mapping}\ }\textbf {\bibinfo {volume} {39}},\ \bibinfo {pages} {1995} (\bibinfo {year} {2018}{\natexlab{b}})}\BibitemShut {NoStop}%
\bibitem [{\citenamefont {Legon}\ \emph {et~al.}(2018{\natexlab{c}})\citenamefont {Legon}, \citenamefont {Bansal}, \citenamefont {Tyshynsky}, \citenamefont {Ai},\ and\ \citenamefont {Mueller}}]{BBB3}%
  \BibitemOpen
  \bibfield  {author} {\bibinfo {author} {\bibfnamefont {W.}~\bibnamefont {Legon}}, \bibinfo {author} {\bibfnamefont {P.}~\bibnamefont {Bansal}}, \bibinfo {author} {\bibfnamefont {R.}~\bibnamefont {Tyshynsky}}, \bibinfo {author} {\bibfnamefont {L.}~\bibnamefont {Ai}}, \ and\ \bibinfo {author} {\bibfnamefont {J.~K.}\ \bibnamefont {Mueller}},\ }\href@noop {} {\bibfield  {journal} {\bibinfo  {journal} {Scientific reports}\ }\textbf {\bibinfo {volume} {8}},\ \bibinfo {pages} {10007} (\bibinfo {year} {2018}{\natexlab{c}})}\BibitemShut {NoStop}%
\bibitem [{\citenamefont {Strohman}\ \emph {et~al.}(2024)\citenamefont {Strohman}, \citenamefont {Payne}, \citenamefont {In}, \citenamefont {Stebbins},\ and\ \citenamefont {Legon}}]{BBB4}%
  \BibitemOpen
  \bibfield  {author} {\bibinfo {author} {\bibfnamefont {A.}~\bibnamefont {Strohman}}, \bibinfo {author} {\bibfnamefont {B.}~\bibnamefont {Payne}}, \bibinfo {author} {\bibfnamefont {A.}~\bibnamefont {In}}, \bibinfo {author} {\bibfnamefont {K.}~\bibnamefont {Stebbins}}, \ and\ \bibinfo {author} {\bibfnamefont {W.}~\bibnamefont {Legon}},\ }\href@noop {} {\bibfield  {journal} {\bibinfo  {journal} {Journal of Neuroscience}\ } (\bibinfo {year} {2024})}\BibitemShut {NoStop}%
\bibitem [{\citenamefont {Lipsman}\ \emph {et~al.}(2013)\citenamefont {Lipsman}, \citenamefont {Schwartz}, \citenamefont {Huang}, \citenamefont {Lee}, \citenamefont {Sankar}, \citenamefont {Chapman}, \citenamefont {Hynynen},\ and\ \citenamefont {Lozano}}]{ablation1}%
  \BibitemOpen
  \bibfield  {author} {\bibinfo {author} {\bibfnamefont {N.}~\bibnamefont {Lipsman}}, \bibinfo {author} {\bibfnamefont {M.~L.}\ \bibnamefont {Schwartz}}, \bibinfo {author} {\bibfnamefont {Y.}~\bibnamefont {Huang}}, \bibinfo {author} {\bibfnamefont {L.}~\bibnamefont {Lee}}, \bibinfo {author} {\bibfnamefont {T.}~\bibnamefont {Sankar}}, \bibinfo {author} {\bibfnamefont {M.}~\bibnamefont {Chapman}}, \bibinfo {author} {\bibfnamefont {K.}~\bibnamefont {Hynynen}}, \ and\ \bibinfo {author} {\bibfnamefont {A.~M.}\ \bibnamefont {Lozano}},\ }\href@noop {} {\bibfield  {journal} {\bibinfo  {journal} {The Lancet Neurology}\ }\textbf {\bibinfo {volume} {12}},\ \bibinfo {pages} {462} (\bibinfo {year} {2013})}\BibitemShut {NoStop}%
\bibitem [{\citenamefont {Xu}\ \emph {et~al.}(2021)\citenamefont {Xu}, \citenamefont {Hall}, \citenamefont {Vlaisavljevich},\ and\ \citenamefont {Lee~Jr}}]{histo1}%
  \BibitemOpen
  \bibfield  {author} {\bibinfo {author} {\bibfnamefont {Z.}~\bibnamefont {Xu}}, \bibinfo {author} {\bibfnamefont {T.~L.}\ \bibnamefont {Hall}}, \bibinfo {author} {\bibfnamefont {E.}~\bibnamefont {Vlaisavljevich}}, \ and\ \bibinfo {author} {\bibfnamefont {F.~T.}\ \bibnamefont {Lee~Jr}},\ }\href@noop {} {\bibfield  {journal} {\bibinfo  {journal} {International Journal of Hyperthermia}\ }\textbf {\bibinfo {volume} {38}},\ \bibinfo {pages} {561} (\bibinfo {year} {2021})}\BibitemShut {NoStop}%
\bibitem [{\citenamefont {Kim}\ \emph {et~al.}(2014)\citenamefont {Kim}, \citenamefont {Hall}, \citenamefont {Xu},\ and\ \citenamefont {Cain}}]{histo2}%
  \BibitemOpen
  \bibfield  {author} {\bibinfo {author} {\bibfnamefont {Y.}~\bibnamefont {Kim}}, \bibinfo {author} {\bibfnamefont {T.~L.}\ \bibnamefont {Hall}}, \bibinfo {author} {\bibfnamefont {Z.}~\bibnamefont {Xu}}, \ and\ \bibinfo {author} {\bibfnamefont {C.~A.}\ \bibnamefont {Cain}},\ }\href@noop {} {\bibfield  {journal} {\bibinfo  {journal} {IEEE transactions on ultrasonics, ferroelectrics, and frequency control}\ }\textbf {\bibinfo {volume} {61}},\ \bibinfo {pages} {582} (\bibinfo {year} {2014})}\BibitemShut {NoStop}%
\bibitem [{\citenamefont {Sukovich}\ \emph {et~al.}(2018)\citenamefont {Sukovich}, \citenamefont {Cain}, \citenamefont {Pandey}, \citenamefont {Chaudhary}, \citenamefont {Camelo-Piragua}, \citenamefont {Allen}, \citenamefont {Hall}, \citenamefont {Snell}, \citenamefont {Xu}, \citenamefont {Cannata} \emph {et~al.}}]{histo3}%
  \BibitemOpen
  \bibfield  {author} {\bibinfo {author} {\bibfnamefont {J.~R.}\ \bibnamefont {Sukovich}}, \bibinfo {author} {\bibfnamefont {C.~A.}\ \bibnamefont {Cain}}, \bibinfo {author} {\bibfnamefont {A.~S.}\ \bibnamefont {Pandey}}, \bibinfo {author} {\bibfnamefont {N.}~\bibnamefont {Chaudhary}}, \bibinfo {author} {\bibfnamefont {S.}~\bibnamefont {Camelo-Piragua}}, \bibinfo {author} {\bibfnamefont {S.~P.}\ \bibnamefont {Allen}}, \bibinfo {author} {\bibfnamefont {T.~L.}\ \bibnamefont {Hall}}, \bibinfo {author} {\bibfnamefont {J.}~\bibnamefont {Snell}}, \bibinfo {author} {\bibfnamefont {Z.}~\bibnamefont {Xu}}, \bibinfo {author} {\bibfnamefont {J.~M.}\ \bibnamefont {Cannata}},  \emph {et~al.},\ }\href@noop {} {\bibfield  {journal} {\bibinfo  {journal} {Journal of neurosurgery}\ }\textbf {\bibinfo {volume} {131}},\ \bibinfo {pages} {1331} (\bibinfo {year} {2018})}\BibitemShut {NoStop}%
\bibitem [{\citenamefont {Kyriakou}\ \emph {et~al.}(2014)\citenamefont {Kyriakou}, \citenamefont {Neufeld}, \citenamefont {Werner}, \citenamefont {Paulides}, \citenamefont {Szekely},\ and\ \citenamefont {Kuster}}]{aberrations1}%
  \BibitemOpen
  \bibfield  {author} {\bibinfo {author} {\bibfnamefont {A.}~\bibnamefont {Kyriakou}}, \bibinfo {author} {\bibfnamefont {E.}~\bibnamefont {Neufeld}}, \bibinfo {author} {\bibfnamefont {B.}~\bibnamefont {Werner}}, \bibinfo {author} {\bibfnamefont {M.~M.}\ \bibnamefont {Paulides}}, \bibinfo {author} {\bibfnamefont {G.}~\bibnamefont {Szekely}}, \ and\ \bibinfo {author} {\bibfnamefont {N.}~\bibnamefont {Kuster}},\ }\href@noop {} {\bibfield  {journal} {\bibinfo  {journal} {International journal of hyperthermia}\ }\textbf {\bibinfo {volume} {30}},\ \bibinfo {pages} {36} (\bibinfo {year} {2014})}\BibitemShut {NoStop}%
\bibitem [{\citenamefont {Almquist}, \citenamefont {Parker},\ and\ \citenamefont {Christensen}(2016)}]{aberrations2}%
  \BibitemOpen
  \bibfield  {author} {\bibinfo {author} {\bibfnamefont {S.}~\bibnamefont {Almquist}}, \bibinfo {author} {\bibfnamefont {D.~L.}\ \bibnamefont {Parker}}, \ and\ \bibinfo {author} {\bibfnamefont {D.~A.}\ \bibnamefont {Christensen}},\ }\href@noop {} {\bibfield  {journal} {\bibinfo  {journal} {Journal of therapeutic ultrasound}\ }\textbf {\bibinfo {volume} {4}},\ \bibinfo {pages} {1} (\bibinfo {year} {2016})}\BibitemShut {NoStop}%
\bibitem [{\citenamefont {Maimbourg}\ \emph {et~al.}(2018)\citenamefont {Maimbourg}, \citenamefont {Houdouin}, \citenamefont {Deffieux}, \citenamefont {Tanter},\ and\ \citenamefont {Aubry}}]{holotrans1}%
  \BibitemOpen
  \bibfield  {author} {\bibinfo {author} {\bibfnamefont {G.}~\bibnamefont {Maimbourg}}, \bibinfo {author} {\bibfnamefont {A.}~\bibnamefont {Houdouin}}, \bibinfo {author} {\bibfnamefont {T.}~\bibnamefont {Deffieux}}, \bibinfo {author} {\bibfnamefont {M.}~\bibnamefont {Tanter}}, \ and\ \bibinfo {author} {\bibfnamefont {J.-F.}\ \bibnamefont {Aubry}},\ }\href@noop {} {\bibfield  {journal} {\bibinfo  {journal} {Physics in Medicine \& Biology}\ }\textbf {\bibinfo {volume} {63}},\ \bibinfo {pages} {025026} (\bibinfo {year} {2018})}\BibitemShut {NoStop}%
\bibitem [{\citenamefont {Kook}\ \emph {et~al.}(2023)\citenamefont {Kook}, \citenamefont {Jo}, \citenamefont {Oh}, \citenamefont {Liang}, \citenamefont {Kim}, \citenamefont {Lee}, \citenamefont {Kim}, \citenamefont {Choi},\ and\ \citenamefont {Lee}}]{holotrans2}%
  \BibitemOpen
  \bibfield  {author} {\bibinfo {author} {\bibfnamefont {G.}~\bibnamefont {Kook}}, \bibinfo {author} {\bibfnamefont {Y.}~\bibnamefont {Jo}}, \bibinfo {author} {\bibfnamefont {C.}~\bibnamefont {Oh}}, \bibinfo {author} {\bibfnamefont {X.}~\bibnamefont {Liang}}, \bibinfo {author} {\bibfnamefont {J.}~\bibnamefont {Kim}}, \bibinfo {author} {\bibfnamefont {S.-M.}\ \bibnamefont {Lee}}, \bibinfo {author} {\bibfnamefont {S.}~\bibnamefont {Kim}}, \bibinfo {author} {\bibfnamefont {J.-W.}\ \bibnamefont {Choi}}, \ and\ \bibinfo {author} {\bibfnamefont {H.~J.}\ \bibnamefont {Lee}},\ }\href@noop {} {\bibfield  {journal} {\bibinfo  {journal} {Microsystems \& Nanoengineering}\ }\textbf {\bibinfo {volume} {9}},\ \bibinfo {pages} {45} (\bibinfo {year} {2023})}\BibitemShut {NoStop}%
\bibitem [{\citenamefont {Hu}\ \emph {et~al.}(2022)\citenamefont {Hu}, \citenamefont {Yang}, \citenamefont {Xu}, \citenamefont {Hao},\ and\ \citenamefont {Chen}}]{holotrans3}%
  \BibitemOpen
  \bibfield  {author} {\bibinfo {author} {\bibfnamefont {Z.}~\bibnamefont {Hu}}, \bibinfo {author} {\bibfnamefont {Y.}~\bibnamefont {Yang}}, \bibinfo {author} {\bibfnamefont {L.}~\bibnamefont {Xu}}, \bibinfo {author} {\bibfnamefont {Y.}~\bibnamefont {Hao}}, \ and\ \bibinfo {author} {\bibfnamefont {H.}~\bibnamefont {Chen}},\ }\href@noop {} {\bibfield  {journal} {\bibinfo  {journal} {Frontiers in Neuroscience}\ }\textbf {\bibinfo {volume} {16}},\ \bibinfo {pages} {984953} (\bibinfo {year} {2022})}\BibitemShut {NoStop}%
\bibitem [{\citenamefont {Jim{\'e}nez-Gamb{\'\i}n}\ \emph {et~al.}(2021)\citenamefont {Jim{\'e}nez-Gamb{\'\i}n}, \citenamefont {Jim{\'e}nez}, \citenamefont {Pouliopoulos}, \citenamefont {Benlloch}, \citenamefont {Konofagou},\ and\ \citenamefont {Camarena}}]{mouse}%
  \BibitemOpen
  \bibfield  {author} {\bibinfo {author} {\bibfnamefont {S.}~\bibnamefont {Jim{\'e}nez-Gamb{\'\i}n}}, \bibinfo {author} {\bibfnamefont {N.}~\bibnamefont {Jim{\'e}nez}}, \bibinfo {author} {\bibfnamefont {A.~N.}\ \bibnamefont {Pouliopoulos}}, \bibinfo {author} {\bibfnamefont {J.~M.}\ \bibnamefont {Benlloch}}, \bibinfo {author} {\bibfnamefont {E.~E.}\ \bibnamefont {Konofagou}}, \ and\ \bibinfo {author} {\bibfnamefont {F.}~\bibnamefont {Camarena}},\ }\href@noop {} {\bibfield  {journal} {\bibinfo  {journal} {IEEE Transactions on Biomedical Engineering}\ }\textbf {\bibinfo {volume} {69}},\ \bibinfo {pages} {1359} (\bibinfo {year} {2021})}\BibitemShut {NoStop}%
\bibitem [{\citenamefont {He}\ \emph {et~al.}(2021)\citenamefont {He}, \citenamefont {Wu}, \citenamefont {Zhu}, \citenamefont {Chen}, \citenamefont {Yuan}, \citenamefont {Zeng},\ and\ \citenamefont {Ji}}]{mouse2}%
  \BibitemOpen
  \bibfield  {author} {\bibinfo {author} {\bibfnamefont {J.}~\bibnamefont {He}}, \bibinfo {author} {\bibfnamefont {J.}~\bibnamefont {Wu}}, \bibinfo {author} {\bibfnamefont {Y.}~\bibnamefont {Zhu}}, \bibinfo {author} {\bibfnamefont {Y.}~\bibnamefont {Chen}}, \bibinfo {author} {\bibfnamefont {M.}~\bibnamefont {Yuan}}, \bibinfo {author} {\bibfnamefont {L.}~\bibnamefont {Zeng}}, \ and\ \bibinfo {author} {\bibfnamefont {X.}~\bibnamefont {Ji}},\ }\href@noop {} {\bibfield  {journal} {\bibinfo  {journal} {IEEE Transactions on Ultrasonics, Ferroelectrics, and Frequency Control}\ }\textbf {\bibinfo {volume} {69}},\ \bibinfo {pages} {662} (\bibinfo {year} {2021})}\BibitemShut {NoStop}%
\bibitem [{\citenamefont {Jim{\'e}nez-Gamb{\'\i}n}\ \emph {et~al.}(2019)\citenamefont {Jim{\'e}nez-Gamb{\'\i}n}, \citenamefont {Jim{\'e}nez}, \citenamefont {Benlloch},\ and\ \citenamefont {Camarena}}]{skull}%
  \BibitemOpen
  \bibfield  {author} {\bibinfo {author} {\bibfnamefont {S.}~\bibnamefont {Jim{\'e}nez-Gamb{\'\i}n}}, \bibinfo {author} {\bibfnamefont {N.}~\bibnamefont {Jim{\'e}nez}}, \bibinfo {author} {\bibfnamefont {J.~M.}\ \bibnamefont {Benlloch}}, \ and\ \bibinfo {author} {\bibfnamefont {F.}~\bibnamefont {Camarena}},\ }\href@noop {} {\bibfield  {journal} {\bibinfo  {journal} {Physical Review Applied}\ }\textbf {\bibinfo {volume} {12}},\ \bibinfo {pages} {014016} (\bibinfo {year} {2019})}\BibitemShut {NoStop}%
\bibitem [{\citenamefont {Jim{\'e}nez-Gamb{\'\i}n}, \citenamefont {Jim{\'e}nez},\ and\ \citenamefont {Camarena}(2020)}]{vortices}%
  \BibitemOpen
  \bibfield  {author} {\bibinfo {author} {\bibfnamefont {S.}~\bibnamefont {Jim{\'e}nez-Gamb{\'\i}n}}, \bibinfo {author} {\bibfnamefont {N.}~\bibnamefont {Jim{\'e}nez}}, \ and\ \bibinfo {author} {\bibfnamefont {F.}~\bibnamefont {Camarena}},\ }\href@noop {} {\bibfield  {journal} {\bibinfo  {journal} {Physical Review Applied}\ }\textbf {\bibinfo {volume} {14}},\ \bibinfo {pages} {054070} (\bibinfo {year} {2020})}\BibitemShut {NoStop}%
\bibitem [{\citenamefont {Sallam}\ \emph {et~al.}(2021)\citenamefont {Sallam}, \citenamefont {Meesala}, \citenamefont {Hajj},\ and\ \citenamefont {Shahab}}]{sallamreflective}%
  \BibitemOpen
  \bibfield  {author} {\bibinfo {author} {\bibfnamefont {A.}~\bibnamefont {Sallam}}, \bibinfo {author} {\bibfnamefont {V.~C.}\ \bibnamefont {Meesala}}, \bibinfo {author} {\bibfnamefont {M.~R.}\ \bibnamefont {Hajj}}, \ and\ \bibinfo {author} {\bibfnamefont {S.}~\bibnamefont {Shahab}},\ }\href {\doibase 10.1063/5.0065489} {\bibfield  {journal} {\bibinfo  {journal} {Applied Physics Letters}\ }\textbf {\bibinfo {volume} {119}},\ \bibinfo {pages} {144101} (\bibinfo {year} {2021})},\ \Eprint {http://arxiv.org/abs/https://pubs.aip.org/aip/apl/article-pdf/doi/10.1063/5.0065489/13194586/144101\_1\_online.pdf} {https://pubs.aip.org/aip/apl/article-pdf/doi/10.1063/5.0065489/13194586/144101\_1\_online.pdf} \BibitemShut {NoStop}%
\bibitem [{\citenamefont {Sallam}, \citenamefont {Meesala},\ and\ \citenamefont {Shahab}(2021)}]{sallamreflective2}%
  \BibitemOpen
  \bibfield  {author} {\bibinfo {author} {\bibfnamefont {A.}~\bibnamefont {Sallam}}, \bibinfo {author} {\bibfnamefont {V.~C.}\ \bibnamefont {Meesala}}, \ and\ \bibinfo {author} {\bibfnamefont {S.}~\bibnamefont {Shahab}},\ }in\ \href@noop {} {\emph {\bibinfo {booktitle} {Active and Passive Smart Structures and Integrated Systems XV}}},\ Vol.\ \bibinfo {volume} {11588}\ (\bibinfo {organization} {SPIE},\ \bibinfo {year} {2021})\ p.\ \bibinfo {pages} {1158807}\BibitemShut {NoStop}%
\bibitem [{\citenamefont {Sallam}\ and\ \citenamefont {Shahab}(2022)}]{sallamNL}%
  \BibitemOpen
  \bibfield  {author} {\bibinfo {author} {\bibfnamefont {A.}~\bibnamefont {Sallam}}\ and\ \bibinfo {author} {\bibfnamefont {S.}~\bibnamefont {Shahab}},\ }\href {\doibase 10.1063/5.0123271} {\bibfield  {journal} {\bibinfo  {journal} {Applied Physics Letters}\ }\textbf {\bibinfo {volume} {121}},\ \bibinfo {pages} {204101} (\bibinfo {year} {2022})},\ \Eprint {http://arxiv.org/abs/https://pubs.aip.org/aip/apl/article-pdf/doi/10.1063/5.0123271/16489147/204101\_1\_online.pdf} {https://pubs.aip.org/aip/apl/article-pdf/doi/10.1063/5.0123271/16489147/204101\_1\_online.pdf} \BibitemShut {NoStop}%
\bibitem [{\citenamefont {Melde}\ \emph {et~al.}(2016)\citenamefont {Melde}, \citenamefont {Mark}, \citenamefont {Qiu},\ and\ \citenamefont {Fischer}}]{melde}%
  \BibitemOpen
  \bibfield  {author} {\bibinfo {author} {\bibfnamefont {K.}~\bibnamefont {Melde}}, \bibinfo {author} {\bibfnamefont {A.~G.}\ \bibnamefont {Mark}}, \bibinfo {author} {\bibfnamefont {T.}~\bibnamefont {Qiu}}, \ and\ \bibinfo {author} {\bibfnamefont {P.}~\bibnamefont {Fischer}},\ }\href@noop {} {\bibfield  {journal} {\bibinfo  {journal} {Nature}\ }\textbf {\bibinfo {volume} {537}},\ \bibinfo {pages} {518} (\bibinfo {year} {2016})}\BibitemShut {NoStop}%
\bibitem [{\citenamefont {Bakhtiari-Nejad}\ \emph {et~al.}(2018)\citenamefont {Bakhtiari-Nejad}, \citenamefont {Elnahhas}, \citenamefont {Hajj},\ and\ \citenamefont {Shahab}}]{marjan}%
  \BibitemOpen
  \bibfield  {author} {\bibinfo {author} {\bibfnamefont {M.}~\bibnamefont {Bakhtiari-Nejad}}, \bibinfo {author} {\bibfnamefont {A.}~\bibnamefont {Elnahhas}}, \bibinfo {author} {\bibfnamefont {M.~R.}\ \bibnamefont {Hajj}}, \ and\ \bibinfo {author} {\bibfnamefont {S.}~\bibnamefont {Shahab}},\ }\href@noop {} {\bibfield  {journal} {\bibinfo  {journal} {Journal of Applied Physics}\ }\textbf {\bibinfo {volume} {124}} (\bibinfo {year} {2018})}\BibitemShut {NoStop}%
\bibitem [{\citenamefont {Melde}\ \emph {et~al.}(2023)\citenamefont {Melde}, \citenamefont {Kremer}, \citenamefont {Shi}, \citenamefont {Seneca}, \citenamefont {Frey}, \citenamefont {Platzman}, \citenamefont {Degel}, \citenamefont {Schmitt}, \citenamefont {Sch{\"o}lkopf},\ and\ \citenamefont {Fischer}}]{volumetric}%
  \BibitemOpen
  \bibfield  {author} {\bibinfo {author} {\bibfnamefont {K.}~\bibnamefont {Melde}}, \bibinfo {author} {\bibfnamefont {H.}~\bibnamefont {Kremer}}, \bibinfo {author} {\bibfnamefont {M.}~\bibnamefont {Shi}}, \bibinfo {author} {\bibfnamefont {S.}~\bibnamefont {Seneca}}, \bibinfo {author} {\bibfnamefont {C.}~\bibnamefont {Frey}}, \bibinfo {author} {\bibfnamefont {I.}~\bibnamefont {Platzman}}, \bibinfo {author} {\bibfnamefont {C.}~\bibnamefont {Degel}}, \bibinfo {author} {\bibfnamefont {D.}~\bibnamefont {Schmitt}}, \bibinfo {author} {\bibfnamefont {B.}~\bibnamefont {Sch{\"o}lkopf}}, \ and\ \bibinfo {author} {\bibfnamefont {P.}~\bibnamefont {Fischer}},\ }\href@noop {} {\bibfield  {journal} {\bibinfo  {journal} {Science Advances}\ }\textbf {\bibinfo {volume} {9}},\ \bibinfo {pages} {eadf6182} (\bibinfo {year} {2023})}\BibitemShut {NoStop}%
\bibitem [{\citenamefont {Kim}\ \emph {et~al.}(2021)\citenamefont {Kim}, \citenamefont {Kasoji}, \citenamefont {Durham},\ and\ \citenamefont {Dayton}}]{cavitation}%
  \BibitemOpen
  \bibfield  {author} {\bibinfo {author} {\bibfnamefont {J.}~\bibnamefont {Kim}}, \bibinfo {author} {\bibfnamefont {S.}~\bibnamefont {Kasoji}}, \bibinfo {author} {\bibfnamefont {P.~G.}\ \bibnamefont {Durham}}, \ and\ \bibinfo {author} {\bibfnamefont {P.~A.}\ \bibnamefont {Dayton}},\ }\href@noop {} {\bibfield  {journal} {\bibinfo  {journal} {Applied Physics Letters}\ }\textbf {\bibinfo {volume} {118}} (\bibinfo {year} {2021})}\BibitemShut {NoStop}%
\bibitem [{\citenamefont {Andr{\'e}s}\ \emph {et~al.}(2022{\natexlab{a}})\citenamefont {Andr{\'e}s}, \citenamefont {Vappou}, \citenamefont {Jim{\'e}nez},\ and\ \citenamefont {Camarena}}]{thermal}%
  \BibitemOpen
  \bibfield  {author} {\bibinfo {author} {\bibfnamefont {D.}~\bibnamefont {Andr{\'e}s}}, \bibinfo {author} {\bibfnamefont {J.}~\bibnamefont {Vappou}}, \bibinfo {author} {\bibfnamefont {N.}~\bibnamefont {Jim{\'e}nez}}, \ and\ \bibinfo {author} {\bibfnamefont {F.}~\bibnamefont {Camarena}},\ }\href@noop {} {\bibfield  {journal} {\bibinfo  {journal} {Applied Physics Letters}\ }\textbf {\bibinfo {volume} {120}} (\bibinfo {year} {2022}{\natexlab{a}})}\BibitemShut {NoStop}%
\bibitem [{\citenamefont {Treeby}\ and\ \citenamefont {Cox}(2010)}]{k-wave}%
  \BibitemOpen
  \bibfield  {author} {\bibinfo {author} {\bibfnamefont {B.~E.}\ \bibnamefont {Treeby}}\ and\ \bibinfo {author} {\bibfnamefont {B.~T.}\ \bibnamefont {Cox}},\ }\href@noop {} {\bibfield  {journal} {\bibinfo  {journal} {Journal of biomedical optics}\ }\textbf {\bibinfo {volume} {15}},\ \bibinfo {pages} {021314} (\bibinfo {year} {2010})}\BibitemShut {NoStop}%
\bibitem [{\citenamefont {Sallam}\ and\ \citenamefont {Shahab}(2023)}]{adaptive}%
  \BibitemOpen
  \bibfield  {author} {\bibinfo {author} {\bibfnamefont {A.}~\bibnamefont {Sallam}}\ and\ \bibinfo {author} {\bibfnamefont {S.}~\bibnamefont {Shahab}},\ }\href {\doibase 10.1109/TUFFC.2023.3315011} {\bibfield  {journal} {\bibinfo  {journal} {IEEE Transactions on Ultrasonics, Ferroelectrics, and Frequency Control}\ }\textbf {\bibinfo {volume} {70}},\ \bibinfo {pages} {1516} (\bibinfo {year} {2023})}\BibitemShut {NoStop}%
\bibitem [{\citenamefont {Gu}\ and\ \citenamefont {Jing}(2020)}]{strongly}%
  \BibitemOpen
  \bibfield  {author} {\bibinfo {author} {\bibfnamefont {J.}~\bibnamefont {Gu}}\ and\ \bibinfo {author} {\bibfnamefont {Y.}~\bibnamefont {Jing}},\ }\href@noop {} {\bibfield  {journal} {\bibinfo  {journal} {The Journal of the Acoustical Society of America}\ }\textbf {\bibinfo {volume} {147}},\ \bibinfo {pages} {4055} (\bibinfo {year} {2020})}\BibitemShut {NoStop}%
\bibitem [{\citenamefont {Gu}\ and\ \citenamefont {Jing}(2018)}]{weakly}%
  \BibitemOpen
  \bibfield  {author} {\bibinfo {author} {\bibfnamefont {J.}~\bibnamefont {Gu}}\ and\ \bibinfo {author} {\bibfnamefont {Y.}~\bibnamefont {Jing}},\ }\href@noop {} {\bibfield  {journal} {\bibinfo  {journal} {IEEE transactions on ultrasonics, ferroelectrics, and frequency control}\ }\textbf {\bibinfo {volume} {65}},\ \bibinfo {pages} {1258} (\bibinfo {year} {2018})}\BibitemShut {NoStop}%
\bibitem [{\citenamefont {Gu}\ and\ \citenamefont {Jing}(2021)}]{msound}%
  \BibitemOpen
  \bibfield  {author} {\bibinfo {author} {\bibfnamefont {J.}~\bibnamefont {Gu}}\ and\ \bibinfo {author} {\bibfnamefont {Y.}~\bibnamefont {Jing}},\ }\href@noop {} {\bibfield  {journal} {\bibinfo  {journal} {IEEE transactions on ultrasonics, ferroelectrics, and frequency control}\ }\textbf {\bibinfo {volume} {68}},\ \bibinfo {pages} {1476} (\bibinfo {year} {2021})}\BibitemShut {NoStop}%
\bibitem [{\citenamefont {Zhang}\ \emph {et~al.}(2017)\citenamefont {Zhang}, \citenamefont {P{\'e}gard}, \citenamefont {Zhong}, \citenamefont {Adesnik},\ and\ \citenamefont {Waller}}]{nonconvex}%
  \BibitemOpen
  \bibfield  {author} {\bibinfo {author} {\bibfnamefont {J.}~\bibnamefont {Zhang}}, \bibinfo {author} {\bibfnamefont {N.}~\bibnamefont {P{\'e}gard}}, \bibinfo {author} {\bibfnamefont {J.}~\bibnamefont {Zhong}}, \bibinfo {author} {\bibfnamefont {H.}~\bibnamefont {Adesnik}}, \ and\ \bibinfo {author} {\bibfnamefont {L.}~\bibnamefont {Waller}},\ }\href@noop {} {\bibfield  {journal} {\bibinfo  {journal} {Optica}\ }\textbf {\bibinfo {volume} {4}},\ \bibinfo {pages} {1306} (\bibinfo {year} {2017})}\BibitemShut {NoStop}%
\bibitem [{\citenamefont {Craig}(2009)}]{insula1}%
  \BibitemOpen
  \bibfield  {author} {\bibinfo {author} {\bibfnamefont {A.~D.}\ \bibnamefont {Craig}},\ }\href@noop {} {\bibfield  {journal} {\bibinfo  {journal} {Nature reviews neuroscience}\ }\textbf {\bibinfo {volume} {10}},\ \bibinfo {pages} {59} (\bibinfo {year} {2009})}\BibitemShut {NoStop}%
\bibitem [{\citenamefont {Uddin}(2015)}]{insula2}%
  \BibitemOpen
  \bibfield  {author} {\bibinfo {author} {\bibfnamefont {L.~Q.}\ \bibnamefont {Uddin}},\ }\href@noop {} {\bibfield  {journal} {\bibinfo  {journal} {Nature reviews neuroscience}\ }\textbf {\bibinfo {volume} {16}},\ \bibinfo {pages} {55} (\bibinfo {year} {2015})}\BibitemShut {NoStop}%
\bibitem [{\citenamefont {Eichenbaum}, \citenamefont {Otto},\ and\ \citenamefont {Cohen}(1992)}]{hippocampus}%
  \BibitemOpen
  \bibfield  {author} {\bibinfo {author} {\bibfnamefont {H.}~\bibnamefont {Eichenbaum}}, \bibinfo {author} {\bibfnamefont {T.}~\bibnamefont {Otto}}, \ and\ \bibinfo {author} {\bibfnamefont {N.~J.}\ \bibnamefont {Cohen}},\ }\href@noop {} {\bibfield  {journal} {\bibinfo  {journal} {Behavioral and neural biology}\ }\textbf {\bibinfo {volume} {57}},\ \bibinfo {pages} {2} (\bibinfo {year} {1992})}\BibitemShut {NoStop}%
\bibitem [{\citenamefont {Grahn}, \citenamefont {Parkinson},\ and\ \citenamefont {Owen}(2008)}]{Caudate}%
  \BibitemOpen
  \bibfield  {author} {\bibinfo {author} {\bibfnamefont {J.~A.}\ \bibnamefont {Grahn}}, \bibinfo {author} {\bibfnamefont {J.~A.}\ \bibnamefont {Parkinson}}, \ and\ \bibinfo {author} {\bibfnamefont {A.~M.}\ \bibnamefont {Owen}},\ }\href@noop {} {\bibfield  {journal} {\bibinfo  {journal} {Progress in neurobiology}\ }\textbf {\bibinfo {volume} {86}},\ \bibinfo {pages} {141} (\bibinfo {year} {2008})}\BibitemShut {NoStop}%
\bibitem [{\citenamefont {Janak}\ and\ \citenamefont {Tye}(2015)}]{Amygdala}%
  \BibitemOpen
  \bibfield  {author} {\bibinfo {author} {\bibfnamefont {P.~H.}\ \bibnamefont {Janak}}\ and\ \bibinfo {author} {\bibfnamefont {K.~M.}\ \bibnamefont {Tye}},\ }\href@noop {} {\bibfield  {journal} {\bibinfo  {journal} {Nature}\ }\textbf {\bibinfo {volume} {517}},\ \bibinfo {pages} {284} (\bibinfo {year} {2015})}\BibitemShut {NoStop}%
\bibitem [{\citenamefont {Andr{\'e}s}\ \emph {et~al.}(2022{\natexlab{b}})\citenamefont {Andr{\'e}s}, \citenamefont {Jim{\'e}nez}, \citenamefont {Benlloch},\ and\ \citenamefont {Camarena}}]{numerical}%
  \BibitemOpen
  \bibfield  {author} {\bibinfo {author} {\bibfnamefont {D.}~\bibnamefont {Andr{\'e}s}}, \bibinfo {author} {\bibfnamefont {N.}~\bibnamefont {Jim{\'e}nez}}, \bibinfo {author} {\bibfnamefont {J.~M.}\ \bibnamefont {Benlloch}}, \ and\ \bibinfo {author} {\bibfnamefont {F.}~\bibnamefont {Camarena}},\ }\href@noop {} {\bibfield  {journal} {\bibinfo  {journal} {Ultrasound in Medicine \& Biology}\ }\textbf {\bibinfo {volume} {48}},\ \bibinfo {pages} {872} (\bibinfo {year} {2022}{\natexlab{b}})}\BibitemShut {NoStop}%
\bibitem [{\citenamefont {Antoniou}\ \emph {et~al.}(2021)\citenamefont {Antoniou}, \citenamefont {Evripidou}, \citenamefont {Giannakou}, \citenamefont {Constantinides},\ and\ \citenamefont {Damianou}}]{thermoplastics}%
  \BibitemOpen
  \bibfield  {author} {\bibinfo {author} {\bibfnamefont {A.}~\bibnamefont {Antoniou}}, \bibinfo {author} {\bibfnamefont {N.}~\bibnamefont {Evripidou}}, \bibinfo {author} {\bibfnamefont {M.}~\bibnamefont {Giannakou}}, \bibinfo {author} {\bibfnamefont {G.}~\bibnamefont {Constantinides}}, \ and\ \bibinfo {author} {\bibfnamefont {C.}~\bibnamefont {Damianou}},\ }\href@noop {} {\bibfield  {journal} {\bibinfo  {journal} {The Journal of the Acoustical Society of America}\ }\textbf {\bibinfo {volume} {149}},\ \bibinfo {pages} {2854} (\bibinfo {year} {2021})}\BibitemShut {NoStop}%
\bibitem [{\citenamefont {Bakaric}\ \emph {et~al.}(2021)\citenamefont {Bakaric}, \citenamefont {Miloro}, \citenamefont {Javaherian}, \citenamefont {Cox}, \citenamefont {Treeby},\ and\ \citenamefont {Brown}}]{photopolymers}%
  \BibitemOpen
  \bibfield  {author} {\bibinfo {author} {\bibfnamefont {M.}~\bibnamefont {Bakaric}}, \bibinfo {author} {\bibfnamefont {P.}~\bibnamefont {Miloro}}, \bibinfo {author} {\bibfnamefont {A.}~\bibnamefont {Javaherian}}, \bibinfo {author} {\bibfnamefont {B.~T.}\ \bibnamefont {Cox}}, \bibinfo {author} {\bibfnamefont {B.~E.}\ \bibnamefont {Treeby}}, \ and\ \bibinfo {author} {\bibfnamefont {M.~D.}\ \bibnamefont {Brown}},\ }\href@noop {} {\bibfield  {journal} {\bibinfo  {journal} {The Journal of the Acoustical Society of America}\ }\textbf {\bibinfo {volume} {150}},\ \bibinfo {pages} {2798} (\bibinfo {year} {2021})}\BibitemShut {NoStop}%
\end{thebibliography}%
\end{document}